\documentclass[%
 aip,
 amsmath,amssymb,
reprint,%
]{revtex4-1}

\usepackage{graphicx}
\usepackage{subcaption} 
\usepackage{dcolumn}
\usepackage{bm}
\usepackage[usenames,dvipsnames]{xcolor}
\usepackage[utf8]{inputenc}
\usepackage[T1]{fontenc}
\usepackage{mathptmx}
\usepackage{etoolbox}
\usepackage{natbib}
\usepackage[hidelinks]{hyperref}
\usepackage{array}
\usepackage{multirow}
\usepackage{ulem}

\makeatletter
\def\@email#1#2{%
 \endgroup
 \patchcmd{\titleblock@produce}
  {\frontmatter@RRAPformat}
  {\frontmatter@RRAPformat{\produce@RRAP{*#1\href{mailto:#2}{#2}}}\frontmatter@RRAPformat}
  {}{}
}%
\makeatother

\newcommand{\eqsplit}[1]{\begin{equation} \begin{split} #1 \end{split}\end{equation}}

\newcommand{\ket}[1]{\left\vert #1\right\rangle}

\newcommand{\occ}{\mathrm{occ}}
\newcommand{\vir}{\mathrm{vir}}

\newcommand{\dchemistry}{Department of Chemistry, Rice University, Houston, TX 77005-1892}
\newcommand{\dphysics}{Department of Physics and Astronomy, Rice University, Houston, TX 77005-1892}

\begin{document}
\preprint{AIP/123-QED}

\title{Selected non-orthogonal configuration interaction with compressed single and double excitations}

\author{Chong Sun}
\affiliation{\dchemistry}
\author{Fei Gao}
\affiliation{\dphysics}
\author{Gustavo E. Scuseria}
\affiliation{\dchemistry}
\affiliation{\dphysics}
\email{guscus@rice.edu}

\date{\today}

\begin{abstract}
Addressing both dynamic and static correlation accurately is a primary goal in electronic structure theory. Non-orthogonal configuration interaction (NOCI) is a versatile tool for treating static correlation, offering chemical insights by combining diverse reference states. Nevertheless, achieving quantitative accuracy requires the inclusion of missing dynamic correlation. This work introduces a framework for compressing orthogonal single and double excitations into an NOCI of a much smaller dimension. This compression is repeated with each Slater determinant in a reference NOCI, resulting in another NOCI that includes all its single and double excitations (NOCISD), effectively recovering the missing dynamic correlations from the reference. This compressed NOCISD is further refined through a selection process using metric and energy tests (SNOCISD). We validate the effectiveness of SNOCISD through its application to the dissociation of the nitrogen molecule and the hole-doped two-dimensional Hubbard model at various interaction strengths.
\end{abstract}

\maketitle

\section{\label{sec:intro}Introduction}
Modern quantum chemistry is increasingly focused on accurate simulations of strongly correlated systems. These systems, where Hartree-Fock falls short, include challenges like excited states, nonequilibrium geometries, and many transition-metal compounds, among others. 
Post-Hartree-Fock (post-HF) methods, such as M\o{}ller–Plesset perturbation theory (MP), truncated configuration interaction (CI), and coupled cluster theory (CC), provide improvement over the HF reference by incorporating dynamic (or weak) correlations from particle-hole excitations. 
However, these methods encounter significant challenges when strong (static or non-dynamic) correlations become significant. 
Addressing static correlation often involves considering multiple reference configurations. The Complete Active Space Self-Consistent Field (CASSCF) method,~\cite{Roos1980TheComplete} incorporating all possible configurations within a chosen active space, was developed to tackle static correlation. However, as the active space size increases, exact solvers such as full configuration interaction (FCI) become intractable, hindering large-scale simulations.

On the other hand, non-orthogonal configuration interaction (NOCI)\cite{Thom2009Hartree, Jimenez2013Multi, Jimenez2013Thesis} offers a compact and versatile alternative to CASSCF without sacrificing accuracy.
In NOCI, each electronic configuration, defined by its unique set of molecular orbitals (MOs), is not necessarily orthogonal to others. 
This "different orbitals for different configurations" approach endows NOCI with great flexibility, requiring fewer configurations to describe sets of orthogonal excitations. Its diabatic nature makes NOCI particularly useful in chemically intuitive studies of phenomena like avoided crossings and conical intersections.~\cite{Barca2014Hartree, Jensen2018Modeling, Mahler2021Orbital} 
Methods that can generate an NOCI wavefunction include symmetry breaking and restoration in projected Hartree-Fock (PHF) theory,~\cite{Smeyers1973Half, Smeyers1974Half, Jimenez2012Projected, Garza2013Capturing, Tsuchimochi2016Configuration} or combining non-unitary Thouless rotations~\cite{Thouless1960NucPhys} over a reference Slater determinant (henceforth, determinant). The Thouless parameters defining the latter strategy can be optimized with either the resonating HF (ResHF) approach~\cite{Fukutome1988Theory} or the few-determinant (FED) method.~\cite{Schmid1989Beyond, Schmid2004On}
ResHF optimizes all determinants simultaneously, whereas FED only optimizes determinants added to the expansion, keeping the previous ones frozen. ResHF often yields better energies, but it requires a more involved optimization process not always worth the effort.~\cite{Jimenez2013Multi} As more determinants are included in the NOCI wavefunction, FCI accuracy is asymptotically approached both with ResHF and FED. However, the energy improvement usually reaches a plateau,~\cite{Jimenez2013Multi} highlighting a saturation point where dynamic correlation is added in a less-than-efficient manner. This observation underscores the challenge we want to address in this work: how can we effectively incorporate dynamic correlation into an NOCI whose static correlation is predominantly included by the initial terms of the expansion? We here propose an innovative strategy. 

Dynamic correlation missing from a given multireference wavefunction can be effectively recovered by including low-rank excitations. 
However, applying these concepts to NOCI wavefunctions, which use different MOs for each configuration, is not straightforward. Efforts to tackle this NOCI challenge include Ten-no's addition of CI and CC,~\cite{Tenno1998Superposition} Burton and Thom's second-order perturbation theory,~\cite{Burton2020Reaching} and Nite and Jimenez-Hoyos' implementation of the CISD expansion for each reference configuration.~\cite{Nite2019Efficient} The latter necessitates a significant truncation to manage computational demands.

This paper introduces a framework for compressing non-orthogonal configuration interaction with single and double-electron excitations (NOCISD) from each term, without brute-force truncation. The key component of our method is a strategy for compressing linear combinations of particle-hole excitations from each determinant into a significantly smaller set of non-orthogonal Slater determinants (NOSD). The method is applicable across all excitation orders. Its efficiency is demonstrated by showing that in the spin-orbital basis, only $2$ NOSD are required for reproducing an expansion with all singly-excited states (CIS), and at most $(2N_{\vir} N_{\occ} + 1)$ NOSD are needed for all doubly-excited states (CID), achieving a square-root reduction in expansion length. After compressing the excited state set, we employ a selection process based on metric and energy criteria to further reduce the NOCI expansion without compromising accuracy. The resulting method is named selected NOCISD (SNOCISD); it provides a versatile and compact approach for incorporating residual dynamic correlation into an NOCI wavefunction. The performance of SNOCISD is here benchmarked on the N$_2$ molecule dissociation curve and the ground state energy of the hole-doped two-dimensional Hubbard model at different interaction regimes. In addition, we analyze the SNOCISD spin-spin correlation function of the two-dimensional case, providing insights into its structure.

\section{Theory}
In this section, we outline the SNOCISD algorithm, which begins by generating a small set of non-orthogonal reference determinants through ResHF or FED optimization. Subsequently, a CISD procedure is applied to each reference determinant, followed by their compression. The following step involves selecting the most important determinants from these compressed sets, 
which are then incorporated into an expanded NOCI. Finally,  a generalized eigenvalue problem is solved with respect to the expanded NOCI. 
Detailed formulae and derivations describing the key steps necessary for implementing the SNOCISD algorithm are presented below.

\subsection{Non-orthogonal configuration interaction\label{sec:noci}}

Mutually non-orthogonal Slater determinants (NOSD) can be constructed through Thouless rotations applied to a reference determinant  $|\Phi_0\rangle$. A new determinant is defined by 
\eqsplit{\label{eq:thouless_rotation}
|\Phi(\mathbf{Z})\rangle = e^{Z}|\Phi_0\rangle, \quad
Z = \sum_{a=1}^{N_{\rm vir}}\sum_{i=1}^{N_{\rm occ}} Z_{ai} E_{a}^{i},
}
where $Z$ represents the Thouless rotation operator, characterized by $N_{\vir}N_{\occ}$ parameters $\mathbf{Z}$. The excitation operator $E_{a}^{i} = b^\dagger_a b_i$ transfers an electron from the $i$th occupied orbital to the $a$th virtual orbital. We use the subscripts "occ" and "vir" to represent occupied and virtual spaces, respectively. Assuming $|\Phi_0\rangle$ is normalized, $|\Phi(\mathbf{Z})\rangle$ maintains intermediate normalization, meaning $\langle \Phi(\mathbf{Z}) | \Phi_0\rangle = 1$. In the limiting case where one parameter $Z_{ai} \gg 1$ and the others are zero, $|\Phi(\mathbf{Z})\rangle = |\Phi_0\rangle + Z_{ai}E_i^a |\Phi_0\rangle$ becomes a singly excited state from $|\Phi_0\rangle$. Higher-order excitations can be similarly obtained when two or more parameters are large and the rest are zero. This is the construction yielding the traditional FCI using a linear combination of NOSD.

In our case, we build  the many-body wavefunction as 
\eqsplit{\label{eq:noci_ansatz}
|\Psi\rangle = \sum_{\mu=0}^{L-1} w_\mu |\Phi_\mu\rangle,
}
where $\mathbf{w} = [w_0, \cdots, w_{L-1}]$ are the coefficients of each $|\Phi_\mu\rangle = |\Phi(\mathbf{Z_\mu})\rangle$.  Typically, $|\Phi_0\rangle$ is the HF solution. Each $|\Phi_\mu\rangle$ is based on a distinct set of molecular orbitals (MOs) differing from those derived from HF.

The coefficients $\mathbf{w}$ are obtained by solving a generalized eigenvalue equation, using the generalized Wick's theorem\cite{Lowdin1955Quantum, Ripka1986Quantum, Balian1969Nonunitary, Chen2023Robust} to compute the matrix elements. The NOCI energy is minimized by optimizing the Thouless parameters $\{\mathbf{Z}_\mu\}$ with the aforementioned ResHF or FED methods.

\subsection{Compression of excited configurations\label{sec:compress}}
We next discuss how a small number of NOSD can represent a linear combination of excited configurations like single and double excitations. We work with spin-orbital excitations, while applications to more specific spin cases are straightforward.
We start by adding a real parameter $t$ to the Thouless rotation in Eq.~\eqref{eq:thouless_rotation},
\eqsplit{\label{eq:touless_w_t}
|\Phi(\mathbf{Z}, t)\rangle = e^{t Z} |\Phi_0\rangle,
}
where $|\Phi_0\rangle$ is the reference configuration. The first derivative of $|\Phi(\mathbf{Z}, t)\rangle$ with respect to $t$ at $t = 0$ is a linear combination of singly-excited determinants,
\eqsplit{
\left.\frac{\partial |\Phi(\mathbf{Z}, t)\rangle }{\partial t} \right\vert_{t=0} = 
Z |\Phi_0\rangle = \sum_{a=1}^{N_{\rm vir}}\sum_{i=1}^{N_{\rm occ}} Z_{ai} |\Phi_i^a\rangle,
} where $|\Phi_i^a\rangle = E_{i}^{a} |\Phi_0\rangle$. This derivative can be written as a two-point central difference with finite step size $2\delta t$
\eqsplit{\label{eq:two_point_diff_singles}
\left.\frac{\partial \left\vert\Phi(\mathbf{Z}, t)\right\rangle }{\partial t} \right\vert_{t=0} 
= 
\lim_{\delta t\rightarrow 0} \frac{\ket{\Phi\left(\mathbf{Z}, \delta t\right)} - \ket{\Phi\left(\mathbf{Z}, -\delta t\right)}}{2\delta t} .
}
Therefore, we can approximate a linear combination of $N_{\occ}N_{\vir}$ singly excited configurations $Z |\Phi_0\rangle$ with only two NOSD
\eqsplit{\label{eq:compress_singles}
|\Psi_{\rm S}\rangle  = Z |\Phi_0\rangle  = \lim_{\delta t\rightarrow 0} \frac{\ket{\Phi\left(\mathbf{Z}, \delta t\right)} - \ket{\Phi\left(\mathbf{Z}, -\delta t\right)}}{2\delta t}.
} 

For double excitations, we write the second derivative of $|\Phi(\mathbf{Z}, t)\rangle$ with respect to $t$
\eqsplit{\label{eq:derivitive_doubles}
{Z}^2|\Phi_0\rangle &=\left.\frac{\partial^2 |\Phi(\mathbf{Z}, t)\rangle }{\partial t^2} \right\vert_{t=0}  \\ 
&= \lim_{\delta t\rightarrow 0} \frac{|\Phi(\mathbf{Z}, 2\delta t)\rangle + |\Phi(\mathbf{Z}, -2\delta t)\rangle - 2|\Phi_0\rangle}{4\delta t^2},
}
where ${Z}^2|\Phi_0\rangle$ is approximated by three determinants: two Thouless-parametrized determinants plus the reference.

A linear combination of double excitations is defined as
\eqsplit{\label{eq:cid_wavefunction}
|\Psi_{\text{D}}\rangle = \sum_{\substack{a, b\in\vir \\ i, j\in\occ}} w_{ji}^{ba} |\Phi_{ji}^{ba}\rangle ,
}
where $w_{ji}^{ba}$ are the coefficients, and  $|\Phi_{ji}^{ba}\rangle = E_{j}^{b}E_{i}^{a} |\Phi_0\rangle$. Assigning $p = (b-1) N_{\occ} + j$ and $q = (a-1) N_{\occ} + i$, we define the matricization of $w_{ji}^{ba}$ as 
\eqsplit{
W_{pq} = w_{ji}^{ba}, 
}
where $\mathbf{W}$ is a Hermitian matrix of size $L\times L$, with $L = N_{\vir}N_{\occ}$, and can be chosen real. Hence, we can diagonalize $\mathbf{W}$ and write
\eqsplit{\label{eq:diag_cid_coeffs}
W_{pq} = \sum_{k = 1} ^{L} \lambda_k U_{kp} U_{kq} = \sum_{k = 1} ^{L} \lambda_k U_{k, bj} U_{k, ai},
}
where $\lambda_k$ is the $k$th eigenvalue, and rows of $\mathbf{U}$ are corresponding eigenvectors. Eq.~\eqref{eq:cid_wavefunction} is rewritten as
\eqsplit{\label{apdx:final_cid_spinless}
|\Psi_{\text{D}}\rangle 
&= 
\sum_{k = 1} ^{L} \lambda_k \left(\sum_{j, b} U_{k, bj}{E}_{j}^{b}\right)\left( \sum_{i, a} U_{k,ai}{E}_{i}^{a}\right)|\Phi_0\rangle \\
&= \sum_{k = 1} ^{L} \lambda_k {Z}_k^2|\Phi_0\rangle.
}
Combined with Eq.~\eqref{eq:derivitive_doubles}, in the limit of $\delta t\rightarrow 0$, we arrive at
\eqsplit{\label{eq:compress_doubles}
|\Psi_{\text{D}}\rangle  = 
\sum_{k = 1} ^{L} \lambda_k 
\frac{|\Phi(\mathbf{Z}_k, 2\delta t)\rangle + |\Phi(\mathbf{Z}_k, -2\delta t)\rangle - 2|\Phi_0\rangle}{4\delta t^2}.
}
Therefore, we can represent $|\Psi_{\text{D}}\rangle$ with at most $(2 N_{\vir}N_{\occ} + 1)$ NOSD, including the reference state $|\Phi_0\rangle$. The compression of double excitations brings a square-root reduction of the determinant count.
In practical applications, introducing a threshold $\lambda_{\min}$ for $|\lambda_k|$ allows exclusion of double excitations with negligible contributions. This means that the actual number of $\mathbf{Z_k}$ matrices can be significantly less than the theoretical maximum of $N_{\vir}N_{\occ}$.

A CISD wavefunction is defined as
\eqsplit{\label{eq:cisd_wf}
\ket{\Psi_{\rm CISD}} = w_0 \ket{\Phi_0}  + \sum_{\substack{a\in\vir \\ i\in\occ}} w_{i}^{a} \ket{\Phi_{i}^{a}} +  \sum_{\substack{a, b\in\vir \\ i, j\in\occ}} w_{ji}^{ba} |\Phi_{ji}^{ba}\rangle,
}
where $w_0$, $\{w_i^a\}$, and $\{w_{ji}^{ba}\}$ are the expansion coefficients for the reference determinant, single excitations, and double excitations, respectively. 

Following Eq.~\eqref{eq:compress_singles} and \eqref{eq:compress_doubles}, we arrive at our non-orthogonal representation of $\ket{\Psi_{\rm CISD}} $ in the limit of $\delta t \rightarrow 0$ as
\eqsplit{\label{eq:compression_cisd}
\ket{\Psi_{\rm CISD}} &=\tilde{w}_0 |\Phi_0\rangle \\
& + \frac{1}{2\delta t}\left(\ket{\Phi( \mathbf{Z}^{(1)}, \delta t)} + \ket{\Phi( \mathbf{Z}^{(1)}, -\delta t)}\right) \\
& + \sum_{k = 1} ^{L} \frac{\lambda_k}{4\delta t^2}
\left(|\Phi( \mathbf{Z}_k^{(2)}, 2\delta t)\rangle + |\Phi(\mathbf{Z}_k^{(2)}, - 2\delta t)\rangle \right),
}
where 
$\tilde{w}_0  = w_0 - \sum_{k = 1} ^{L} \frac{\lambda_k}{2\delta t^2}$. 
The Thouless parameters are given by 
\eqsplit{
Z_{ai}^{(1)} &= w_{i}^a, \\
Z_{k, ai}^{(2)} &= U_{k, ai}.
}

\begin{figure}[h!]
    \centering
    \begin{subfigure}[h]{\linewidth}
    \centering
    \includegraphics[width=\linewidth]{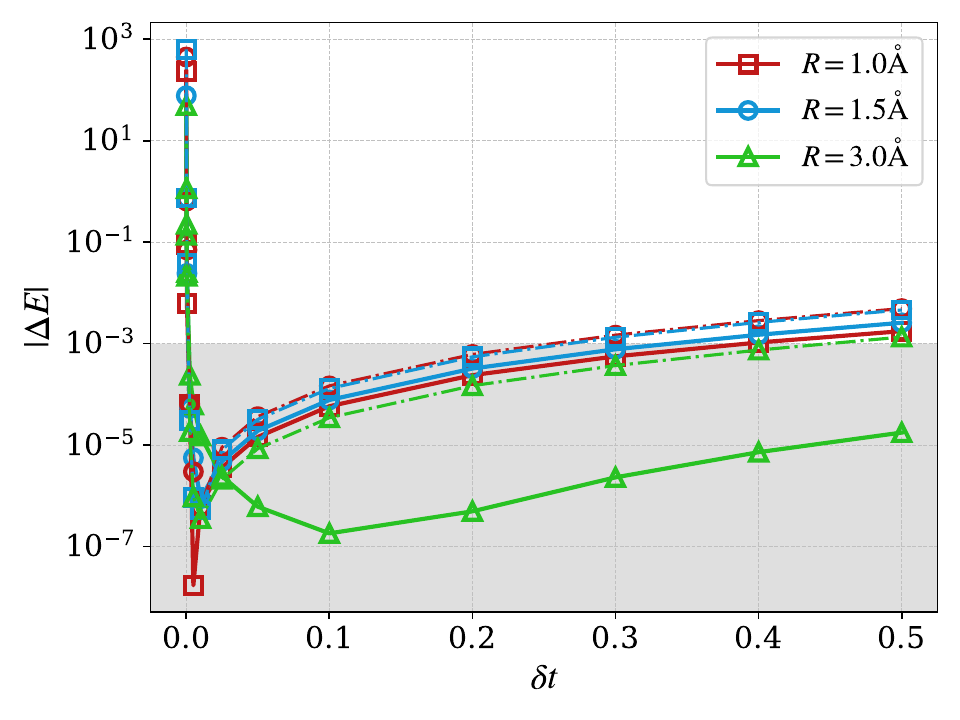}
    \caption{Energy error}\label{fig:cisd_scan_dt_de}
    \end{subfigure}\\
    \begin{subfigure}[h]{\linewidth}
    \centering
    \includegraphics[width=\linewidth]{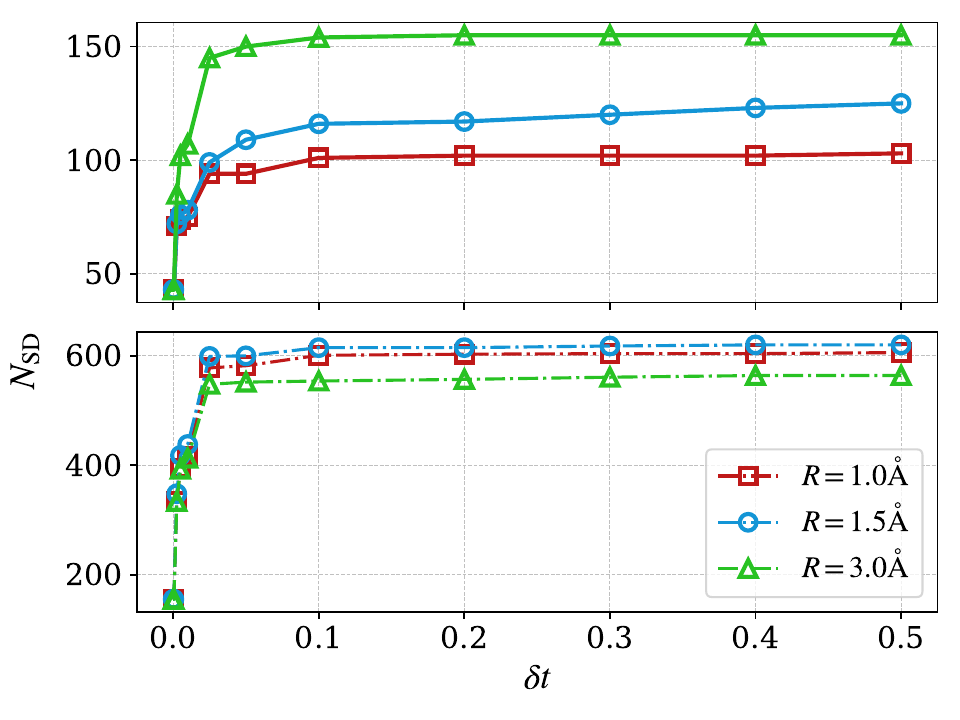}
    \caption{Number of determinants}\label{fig:cisd_scan_dt_ld}
    \end{subfigure}
    \caption{Analysis of UCISD compression as a function of $\delta t$ for N$_2$ with STO-3G (solid lines) and 6-31G (dashed lines) at varying bond lengths ($R$). (a) Absolute value of energy error; (b) Number of linearly independent determinants post-compression. All energies in Hartrees. 
    } 
    \label{fig:cisd_scan_dt}
\end{figure}

In practice, the two-point central difference must be approximated by a nonzero $\delta t$ value, which determines the accuracy of Eq.~\eqref{eq:compression_cisd}.
Fig.~\ref{fig:cisd_scan_dt_de} examines the compression accuracy in unrestricted CISD (UCISD) simulations for the N$_2$ molecule across various $\delta t$ values, employing STO-3G and 6-31G basis sets, and adjusting the bond length to reflect weak to strong correlation regimes.
We observe that the finite difference error decays with decreasing $\delta t$, hitting a point where numerical instability arises. 
An acceptable $\delta t$ range, highlighted in grey shade, corresponds to a compression error below $10^{-3}$ Hartree in the total energy. For most $\delta t$ values between $0$ and $0.5$, the error can be well controlled. Figure~\ref{fig:cisd_scan_dt_ld} illustrates how the number of linearly independent determinants varies with $\delta t$. With an increase in $\delta t$, we observe a rise in the number of determinants, eventually reaching a plateau for the total number of compressed UCISD determinants. 
In the cases examined in this paper, we found that $\delta t = 0.05$ ensures numerical stability and accuracy. Therefore, we have chosen to fix $\delta t = 0.05$ throughout this work.

\begin{figure}[t!]
    \centering
    \begin{subfigure}[h]{\linewidth}
    \centering
    \includegraphics[width=\linewidth]{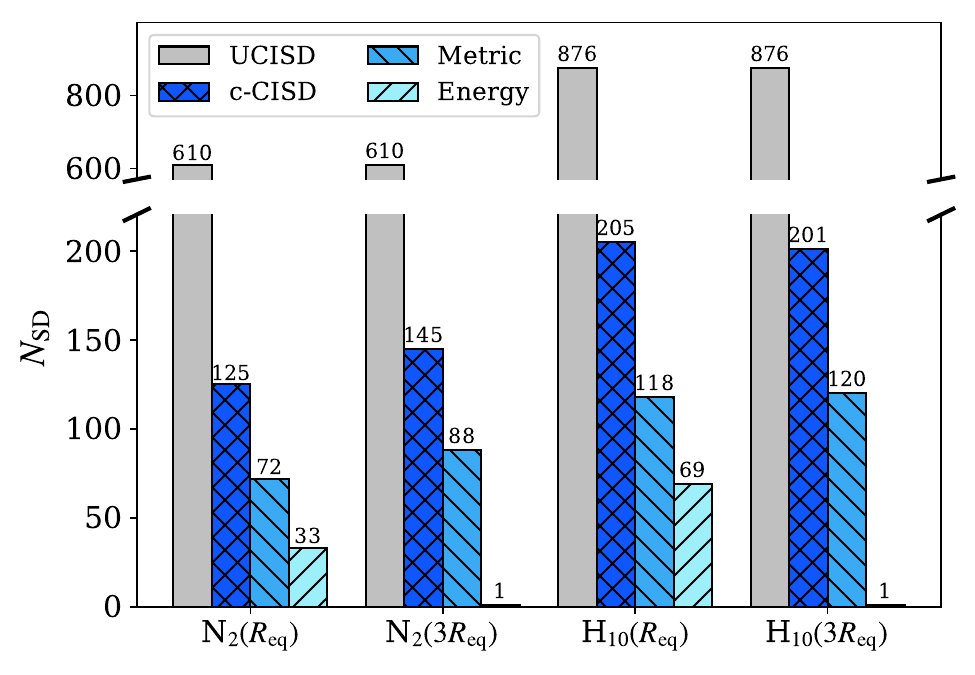}
    \caption{STO-3G}
    \end{subfigure}
    \begin{subfigure}[h]{\linewidth}
    \centering
    \includegraphics[width=\linewidth]{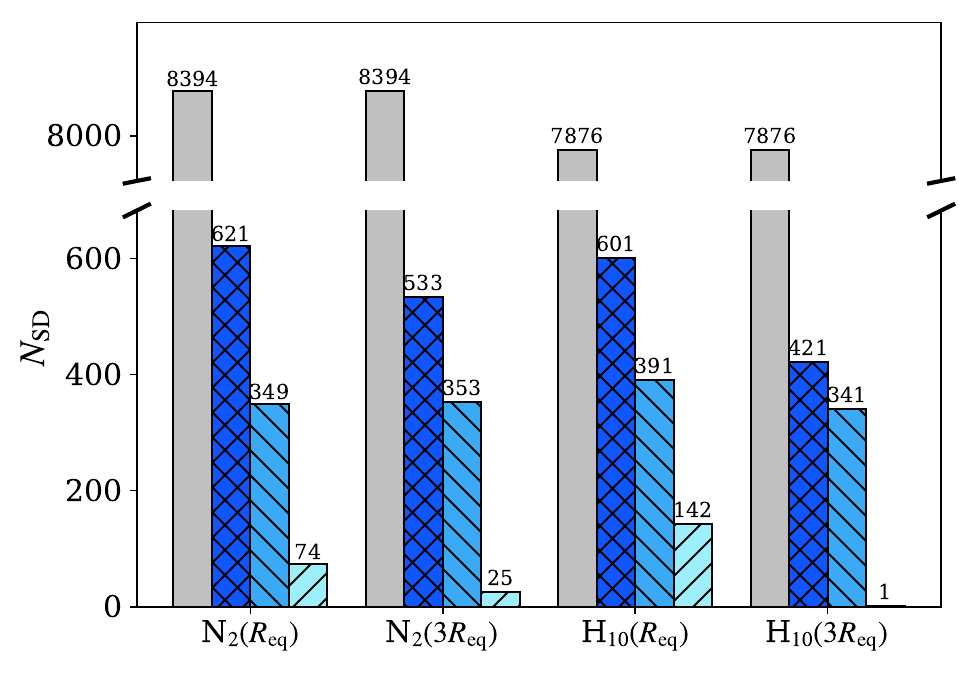}
    \caption{6-31G}
    \end{subfigure}
    \caption{Numbers of determinants of UCISD (grey), after compression (dark blue), after selection based on the metric test (medium blue) and after the energy test (light blue) for N$_2$ and the H$_{10}$ chain using (a) STO-3G and (b) 6-31G basis sets at equilibrium ($R_{\rm eq}$) and extended $3R_{\rm eq}$ bond lengths. 
    } 
    \label{fig:selection}
\end{figure}

\subsection{Selection of NOCISD determinants\label{sec:selection}}
To further refine the NOCISD expansion, we follow a procedure similar to that outlined in work by Dutta et. al.: determinants meeting specific criteria are selected and added to a given set, while others are discarded.~\cite{Dutta2021Construction} We employ a \textit{metric test} with threshold $m_0>0$ and an \textit{energy test} with threshold $h_0>0$.
We use Latin letters $\ket{p}, \ket{q} \cdots$ for determinants in the 
given set $\mathcal{R}$, and Greek letters $\ket{\mu}, \ket{\nu} \cdots$ for determinants to be assessed against the criteria.

We start by defining the projector 
\eqsplit{
{Q} = {I} - \sum_{pq} |p\rangle {X}_{pq} \langle q |,
}
where $\mathbf{X} = \mathbf{M}^{-1}$ and $M_{pq} = \langle p|q\rangle$. ${Q}$ projects a state onto the orthogonal space of $\mathcal{R}$. The metric test requires that 
\eqsplit{\label{eq:select_metric}
\frac{\left\Vert {Q}\ket{\mu}\right\Vert}{\left\Vert \ket{\mu}\right\Vert} \geq m_0.
}

The energy test determines whether a new determinant $|\mu\rangle$ makes a sufficient contribution to the total energy. 
Assuming $E_0$ is the NOCI energy of configurations in $\mathcal{R}$, and $\varepsilon$ is the new NOCI energy after adding $|\mu\rangle$ to $\mathcal{R}$, we require
\eqsplit{\label{eq:select_energy}
\frac{E_0 - \varepsilon}{|E_0|}  > h_0.
}
Details of evaluating $\varepsilon$ are provided in Appendix~\ref{sec:apdx_select}.

\begin{figure}[t!]
    \centering
    \includegraphics[width=\linewidth]{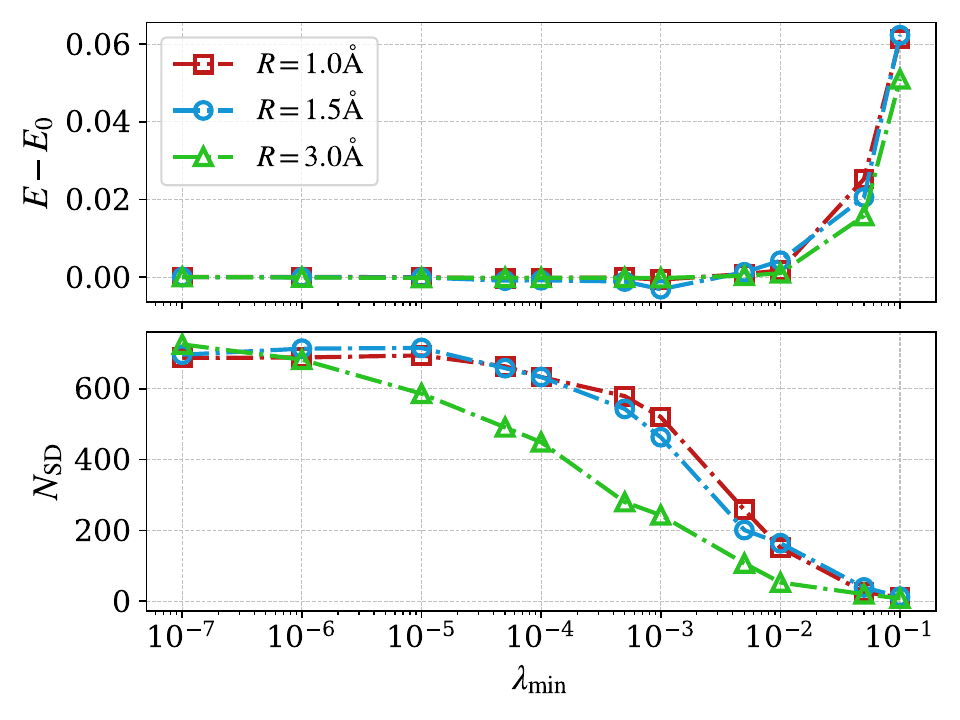}
    \caption{Impact of varying $\lambda_{\min}$ on the ground state energy (in Hartrees) and SNOCISD length for the N$_2$ molecule at various bond lengths. $E_0$ is the energy setting $\lambda_{\min} = 10^{-7}$.}
    \label{fig:scan_tol2}
\end{figure}

Figure~\ref{fig:selection} illustrates the reduction in the number of determinants following UCISD compression and subsequent selection for two systems:  the N$_2$ molecule and a H$_{10}$ chain ($10$ H atoms equally distributed in the $z$ direction), using STO-3G and 6-31G basis sets. 
We chose $\lambda_{\min} = 10^{-7}$, $\delta t = 0.05$, $m_0 = 10^{-5}$ and $h_0 = 10^{-6}$. The equilibrium bond lengths $R_{\rm eq}$ are computed by selected heat bath configuration interaction (SHCI).\cite{Holmes2016Heat, Sharma2017Semistochastic, Li2018Fast}
The reduction in the number of NOSD after compression and selection quite is significant compared to the original UCISD expansion. The reduction is pronounced at larger bond lengths ($3R_{\rm eq}$) regarding the energy test, illustrating the diminishing role of CISD in strongly correlated scenarios.

Practically, given the relatively high computational cost of the energy test, bypassing it can expedite the process. 
Yet, the energy test is essential for minimizing the NOCI expansion in strongly correlated systems, presenting a dilemma: opt for a compact NOCI expansion by conducting the energy test, or avoid it to reduce computational load at the expense of a larger NOCI expansion.
An alternative to removing less important configurations involves stricter truncation of the eigenvalue $\lambda_k$ in Eq.~\eqref{eq:compression_cisd}, i.e., using a larger $\lambda_{\min}$.

Figure~\ref{fig:scan_tol2} shows the impact of varying  $\lambda_{\min}$ on the ground state energy and SNOCISD determinant count for the N$_2$ molecule at various bond lengths within a 6-31G basis set. 
The reference NOCI uses two determinants from a ResHF optimization, setting $\delta t = 0.05$  and $m_0 = 10^{-5}$, skipping the energy test.
Energy differences relative to that with $\lambda_{\min} = 10^{-7}$ are presented in the top panel, while the length of the SNOCISD wavefunction is shown in the bottom panel. 
Notably, the energy starts to increase for $\lambda_{\min} > 10^{-3}$, while the determinant count starts to decrease beyond $\lambda_{\min} > 10^{-5}$, with an even early drop at $R = 3.0$\AA.
This trend identifies an optimal region where the choice of $\lambda_{\min}$ guarantees a compact SNOCISD without compromising the accuracy. 
The shorter SNOCISD expansion at $R = 3.0$\AA\ corroborates with earlier findings that fewer determinants pass the energy test at larger bond lengths.
In the following, we bypass the energy test. 
Instead, meticulous selection of the $\lambda_{\min}$ value ensures a balance between accuracy and efficiency.

\subsection{Computational details}
A homemade Python package is used to carry out the SNOCISD simulations. The optimization in FED and ResHF employed the adaptive moment estimation algorithm (ADAM)~\cite{Kingma2017Adam} implemented in \textsc{Jax}.~\cite{Deepmind2020Jax} The optimization process begins with initial guesses that are near-singly-excited determinants, achieved by 
setting one Thouless parameter much larger than the others ($5:0.1$).
\textsc{PySCF}~\cite{Sun2018PySCF, Sun2020Recent} is used for quantum chemistry tasks such as integral acquisition, HF, CISD, CCSD, and SHCI simulations. \textsc{Block2}~\cite{Zhai2023Block2} is used for density matrix renormalization group (DMRG) calculations.

%
\begin{figure}[t!]
    \centering
    \includegraphics[width=\linewidth]{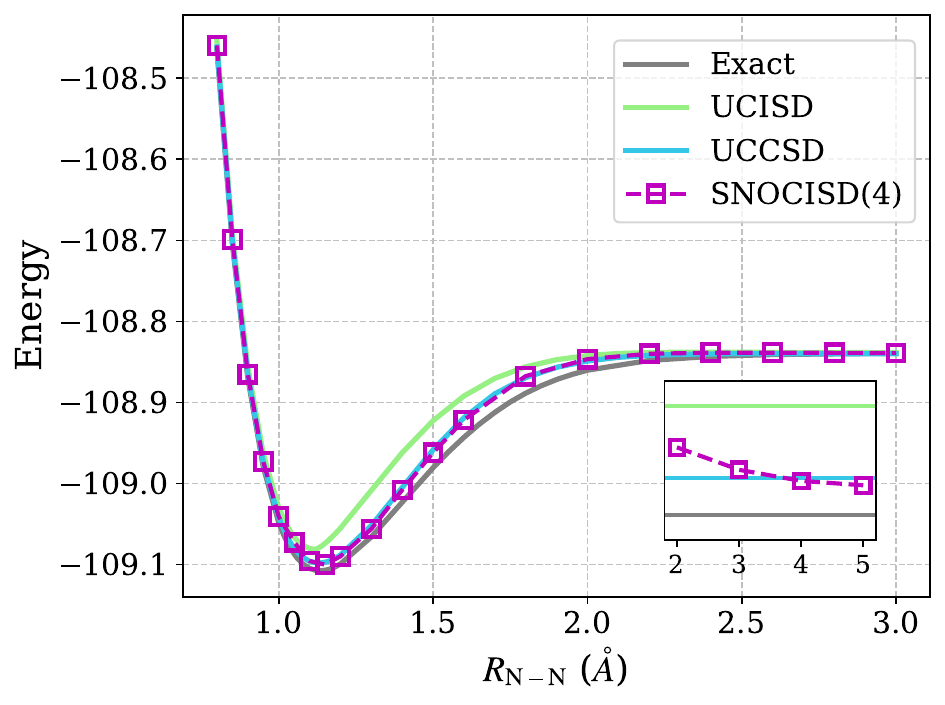}
    \caption{Dissociation energy curve of N$_2$ molecule. Inset: SNOCISD energy with respect to the number of reference determinants at $R_{\mathrm{eq}}$. The energy unit is in Hartree.}
    \label{fig:n2_diss}
\end{figure}
\section{Applications}
\label{sec:results}
\subsection{Dissociation energy\label{sec:results_diss_curve}}
Generating accurate dissociation profiles for molecules is a conventional test for quantum chemistry methods, where systems transition from weak to strong correlations with bond stretching. 
This section assesses SNOCISD by computing the ground-state energy of the N$_2$ molecule across bond lengths from $0.8$ to $3.0$\AA, using the 6-31G basis set, and compares results with SHCI, UCISD, and UCCSD methods. SHCI serves as the benchmark reference, denoted as "Exact" in Fig.~\ref{fig:n2_diss}. A half-spin projection\cite{Mayhall2014SpinFlip} is applied to the SNOCISD wavefunction with $4$ reference configurations generated by FED, shown in Fig.~\ref{fig:n2_diss}. The compression and selection process yields around 350 NOSD for each reference without applying the energy test. We fix $\lambda_{\min} = 10^{-5}$, $m_0 = 10^{-5}$, and $\delta t = 0.05$. 

In the inset plot of Fig.~\ref{fig:n2_diss}, we show the improvement of the SNOCISD energy with respect to the number of references. The $x$-axis represents the number of reference configurations and the $y$-axis the energy at the equilibrium bond length $R_{\rm eq}$. From top to bottom, the horizontal lines correspond to the UCISD, UCCSD and SHCI energies, respectively. We observe that even with $3$ reference configurations, the energy is already comparable to the UCCSD energy. SNOCISD exhibits notable improvements over UCISD and performs comparably to UCCSD, yet it needs further improvement to achieve FCI-level accuracy. One promising direction beyond this work is to include higher-order excitations with careful adaptation of the compression and selection strategies.

\subsection{Two-Dimensional Hubbard model}
\label{sec:2dHub}
The two-dimensional Hubbard model is a challenging system in electronic structure studies. Its scientific significance is underscored by applications in describing the behavior of electrons in real materials, such as Mott insulators and high-temperature superconductors.~\cite{Assaad1996Insulator, Maier2005Systematic, Sordi2010Finite, Gull2013Superconductivity, Zheng2017Stripe}
The Hamiltonian is given by
\eqsplit{
H = -t\sum_{\langle i, j\rangle, \sigma}\left(a_{i, \sigma}^\dagger a_{j, \sigma} + \mathrm{h.c.}\right) + U\sum_{i}n_{i\uparrow}n_{i\downarrow},
}
where $t$ represents the hopping amplitude, and $U > 0$ is the on-site Coulomb interaction strength. $\langle i, j\rangle$ represents nearest-neighbor sites, and $\sigma = \uparrow, \downarrow$ denotes the spin degree of freedom. Conventionally, $t$ is set to $1$, and energies are expressed in units of $t$. As $U$ increases, the system transitions from a weakly correlated to a strongly correlated regime.

\subsubsection{Ground state energy}\label{sec:2dhub_energy}
%
\begin{table}[t!]
\caption{\label{tab:2d_hubbard}Ground state energy per site of the two-dimensional Hubbard model of size $4\times 4$ with periodic boundary conditions (PBC) in both directions. 
}
\begin{tabular}{c|l|cccc}
\hline\hline
  $U/t$    & & 2 & 4 & 8 & 12 \\
\hline
\multirow{ 5}{*}{$f = 0.875$}  & Exact\footnote{DMRG results with bond dimension equal to $2000$.} 
           & -1.1956 & -0.9805 & -0.7377 & -0.6257\\
 & UCISD   & -1.1930 & -0.9558 & -0.6252 & -0.5202\\
 & UCCSD   & -1.1952 & -0.9710 & -0.6717 & -0.5541 \\
 & SNOCISD & -1.1945 & -0.9682 & -0.6896 &- 0.5560\\
 \hline
 \multirow{ 5}{*}{$f = 0.75$} & Exact 
           & -1.2678 & -1.1077 & -0.9314 & -0.8496\\
 & UCISD  & -1.2643 & -1.0811 & -0.7927 & -0.7133\\
 & UCCSD & -1.2663 & -1.0983 & -0.8673 & -0.7644\\
 & SNOCISD & -1.2659 & -1.0932 & -0.8676 & -0.7718 \\
 \hline
\end{tabular}
\end{table}
Accurately evaluating the ground state of the two-dimensional Hubbard model away from half-filling is challenging due to the interplay of competing phases.~\cite{LeBlanc2015Solutions} 
In our study, summarized in Table~\ref{tab:2d_hubbard}, we compute the ground state energy per site for a $4\times 4$ Hubbard lattice under periodic boundary conditions (PBC) in the underdoped regime, comparing SNOCISD with various numerical methods. The number of spin-up and spin-down electrons are set to be the same. 
The SNOCISD simulations used $14$ ResHF-optimized references to address static correlations, and set $\delta t = 0.05$ and $m_0 = 10^{-5}$.
For different $U$ values, slightly different $\lambda_{\min}$ values are picked. In particular, $\lambda_{\min} = 0.001$ is used for $U = 2$ and $4$, while $\lambda_{\min} = 0.005$ for $U = 8$ and $12$. As previously discussed in Section~\ref{sec:selection}, in the strongly correlated scenario, CISD contributes little to the correlation energy. Therefore, it is plausible to introduce a harsher $\lambda_{\min}$ for larger $U$ values.
DMRG results, with a bond dimension of $2000$, serve as our benchmark, denoted as the exact value in Table~\ref{tab:2d_hubbard}.
The SNOCISD method not only shows considerable improvement over UCISD but also outperforms UCCSD at higher $U$ values, showing potential in accurately capturing the ground state energies in complex correlated systems.

\subsubsection{Spin-spin correlation}\label{sec:spin_corr}

The existence of broken spin symmetry suggests the presence of degenerate or near-degenerate spin configurations, where starting from a UHF solution and gradually adding excited configurations reveals a consistent antiferromagnetic (AFM) pattern.
However, achieving the FCI solution eliminates the AFM order, as FCI encompasses all configurations. SNOCISD aims to include the low-lying near-degenerate states, and thus may not precisely capture the long-range AFM order. Nevertheless, employing SNOCISD for magnetic order studies clarifies the contribution of each optimized reference state to the overall wavefunction.
The paramagnetic (PM) and AFM phases can be distinguished by the spin-spin correlation function, $C_{s}(i, j) = \langle n_{i\uparrow}n_{j\downarrow}\rangle - \langle n_{i\uparrow}\rangle\langle n_{j\downarrow}\rangle$, which evaluates the spin-spin correlation (or magnetic correlation) between site-$i$ and site-$j$.
\begin{figure}[t!]
    \centering
    \includegraphics[width=\linewidth]{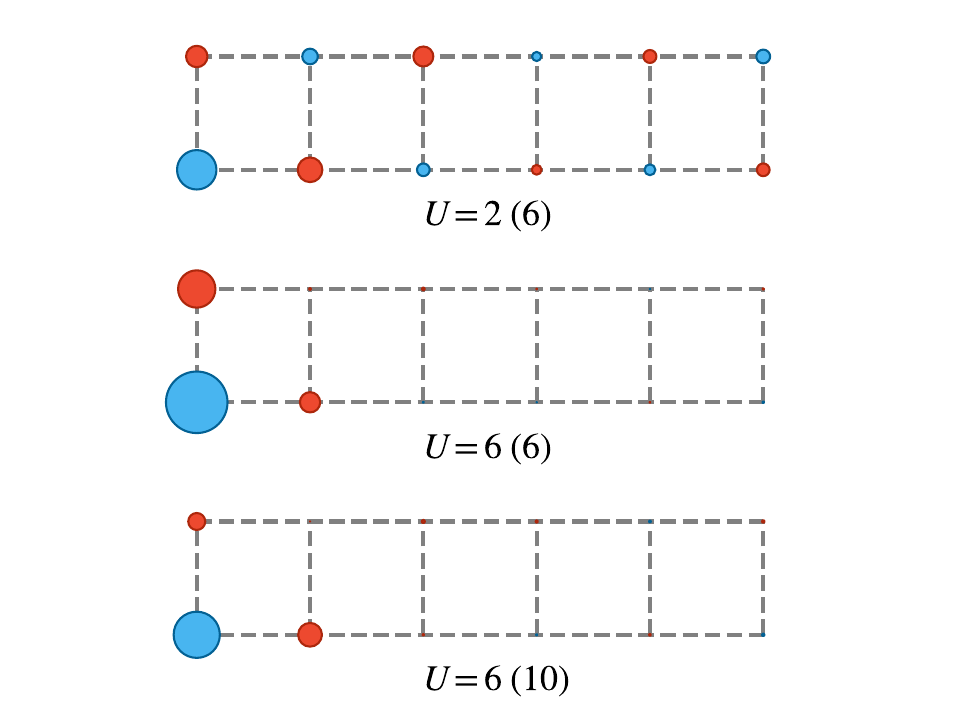}
    \caption{Spin-spin correlation between site-$0$ and site-$i$ ($i = 0, \cdots, 11$) for the two-dimensional Hubbard model of size $2\times 6$ with open boundary conditions (OBC). The numbers in the parentheses represent the number of reference determinants.}
    \label{fig:spin_corr}
\end{figure}

Figure~\ref{fig:spin_corr} presents the SNOCISD-computed spin-spin correlations for a half-filled 8×2 Hubbard model with open boundary conditions (OBC).
$C_{s}(0, i)$ between the lower-left site (denoted as site-$0$) and other sites ($i = 1, \cdots, 11$) is computed.
The SNOCISD wavefunction employs parameters $\lambda_{\min} = 10^{-5}$, $\delta t = 0.05$, and $m_0 = 10^{-5}$. 
Circle radii are proportional to the absolute value of $C_{s}(0, i)$ with blue indicating positive and red negative values. 
The decaying absolute value with respect to the distance from site-$0$ and an alternating sign pattern confirms the AFM order. 
At a lower interaction strength ($U = 2$), a faint AFM order is noticeable, diminishing with distance. However, at a moderate interaction strength ($U = 6$), the AFM pattern is observed only among nearest neighbors and disappears at longer distances.
This pattern diminishes further with an increased number of reference determinants (lowest panel).
At lower $U$ values, with less spin degeneracy, the optimization of reference determinants for NOCI energy captures low excitations from the UHF solution, preserving a weak AFM order. However, at higher $U$ values, the optimization begins to encompass near-degenerate spin configurations, reducing the bias towards any broken-symmetry configuration. This hypothesis is confirmed by the reduced $C_{s}(0, i)$ values as more reference determinants are added.

\section{Conclusions}\label{sec:conclusion}
In this work, we have presented a systematically improvable approach to incorporate dynamic correlation into non-orthogonal configuration interaction (NOCI) wavefunctions, achieved by adding a selected set of singly and doubly excited configurations.
Notably, we prove that any linear combination of orthogonal excited configurations can be accurately represented by a much smaller set of non-orthogonal configurations, leading to significant efficiency gains.
Further refinement is achieved through a selection process based on metric and energy tests. This combined approach, termed selected non-orthogonal configuration interaction with single- and double-electron excitations (SNOCISD), offers an efficient method for simultaneously capturing both dynamic and static correlations.

SNOCISD relies on two key components: compressing the orthogonal CISD wavefunction and selecting configurations via metric and energy tests.
For compression, we used a two-point central difference strategy for approximating orthogonal configurations with Thouless-parameterized non-orthogonal expansions. 
The metric test assesses the overlap of new determinants with the reference set, while the energy test uses a simplified energy evaluation problem. 
We have analyzed the impact of key parameters, including two-point difference step size $\delta t$, the eigenvalue cutoff of double excitations $\lambda_{\min}$, metric threshold $m_0$ and the Hamiltonian threshold $h_0$, on the accuracy and computational cost.
SNOCISD's capabilities are demonstrated by replicating the N$_2$ molecule's energy dissociation curve and studying the ground state energy and magnetic correlation of the two-dimensional Hubbard model.
Compared to UCISD, SNOCISD achieved significant improvements and even surpasses UCCSD in the strong correlation domain.

In summary, SNOCISD provides a promising path towards a computationally efficient and accurate multireference tool. Nonetheless, further development is necessary to improve the method's potential for large-scale simulations and even higher accuracy. Incorporating higher-order excitations and further refining the configuration selection processes are key directions for future studies.

\appendix
\counterwithin{figure}{section}
\section{Selecting determinants based on energy contribution\label{sec:apdx_select}}

\begin{figure}[h!]
    \centering
    \begin{subfigure}[h]{\linewidth}
    \centering
    \includegraphics[width=\textwidth]{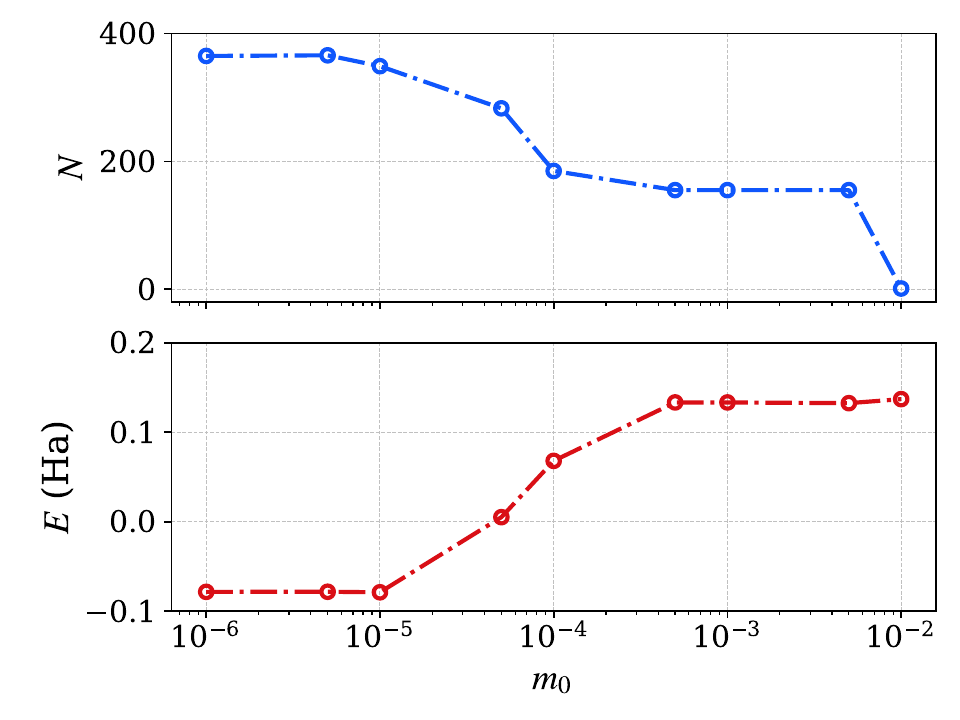}
    \caption{Metric test.}
    \end{subfigure}\\
    \begin{subfigure}[h]{\linewidth}
    \centering
    \includegraphics[width=\textwidth]{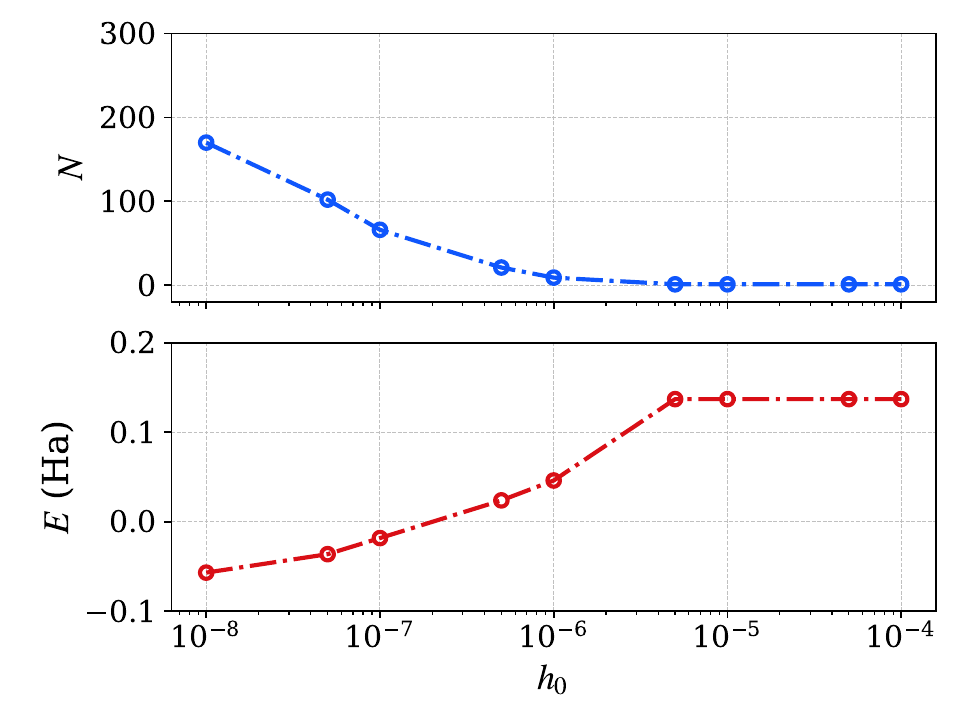}
    \caption{Energy test.}
    \end{subfigure}
    \caption{Number of NOSDs selected from compressed UCISD (blue lines) and NOCI energy (red lines) with respect to the values of (a) $m_0$ and (b) $h_0$. The system simulated is N$_2$ molecule at $R_{\rm eq}$ with 6-31G. The energy is shifted by $-109.0$ Hartree.
    } 
    \label{fig:apdx_error_selection}
\end{figure}
We derive the expression of $\varepsilon$ in the energy test Eq.~\eqref{eq:select_energy}.
We define the solution in the given set $\mathcal{R}$ as
\eqsplit{
|\Psi_0\rangle = \sum_p w_p |p\rangle, \quad E_0 = \frac{\langle \Psi_0 | H |\Psi_0\rangle}{\langle \Psi_0 |\Psi_0\rangle}.
}
Adding ${Q}|\mu\rangle$ to $|\Psi_0\rangle$ results in the wavefunction
\eqsplit{
|\Psi\rangle = \eta_0 |\Psi_0\rangle + \eta_1 {Q}|\mu\rangle.
}
The new ground-state energy and corresponding coefficients $[\eta_0, \eta_1]$ are derived from solving a $2\times 2$ generalized eigenvalue problem 
\eqsplit{\label{eq:apdx_2x2_gev}
\mathbf{Hv} = \varepsilon \mathbf{Sv},
}
where
\eqsplit{ 
\mathbf{S} &= \begin{pmatrix}
    \langle \Psi_0|\Psi_0\rangle & 0 \\ 0 & \langle \mu |{Q}^2|\mu\rangle 
\end{pmatrix} = 
\begin{pmatrix}
     s_0 & 0 \\ 0 & s_\mu
\end{pmatrix}, \\
\mathbf{H} &= \begin{pmatrix}
    \langle \Psi_0| {H}|\Psi_0\rangle & \langle \Psi_0| {H}{Q}|\mu\rangle \\
     \langle \mu |{Q} {H}| \Psi_0\rangle  & \langle \mu|{Q}{H}{Q}|\mu\rangle
\end{pmatrix}
     = \begin{pmatrix}
         H_0 & T \\ T^* & H_{\mu}
     \end{pmatrix}.
}

We require that the new energy after adding ${Q}|\mu\rangle$ passes the energy test
\eqsplit{
\frac{E_0  - \varepsilon}{|E_0 |} > h_0.
}
Here we emphasize that $\varepsilon$ is not the exact NOCI energy of $|\Psi\rangle = \sum_p w_p|p\rangle+ w_\mu |\mu\rangle$, where all coefficients are relaxed and solved simultaneously. However, $\varepsilon$ is a good enough approximation to guide the selection. Moreover, in each selection step, we only need to solve a $2\times 2$ generalized eigenvalue problem, which is much more economical than solving for the exact NOCI energy.

In Fig.~\ref{fig:apdx_error_selection}, we profile the truncation error with respect to the values of $m_0$ and $h_0$. The set of NOSD to be selected is from compression of UCISD for the N$_2$ molecule at $R_{\rm eq}$ with 6-31G. We choose $\delta t = 0.05$ and $\lambda_{\min} = 10^{-7}$. Fig.~\ref{fig:apdx_error_selection}(a) shows the number of SDs and the energy with respect to the value of $m_0$, and it suggests that $m_0 < 10^{-5}$ produces small truncation errors. On the other hand, when $m_0$ is too small, the linear dependency is not fully removed, and singularity can occur when calculating the NOCI energy. Fig.~\ref{fig:apdx_error_selection}(b) shows the number of NOSDs and the energy with respect to the value of $h_0$, where we first removed linear dependency with fixed $m_0 = 10^{-6}$. Understandably, the energy is more sensitive to the value of $h_0$. The error comes from three aspects: i) $\varepsilon$ being not the exact NOCI energy, ii) error accumulation, and iii) correlation among discarded determinants. In practice, one should choose the $h_0$ value according to the desired number of SDs and computing resources.

\begin{acknowledgments}
This work was supported by the U.S. Department of Energy, Office of Basic Energy Sciences, Computational and Theoretical Chemistry Program under Award DE-SC0001474. G.E.S. acknowledges support as a Welch Foundation Chair (Grant No. C-0036). This work was also partly supported by the Big-Data Private-Cloud Research Cyberinfrastructure MRI-award funded by NSF under grant CNS-1338099 and by Rice University's Center for Research Computing (CRC). C.S. would like to thank Austin Cheng, Carlos Jimenez-Hoyos, Ruiheng Song, Thomas Henderson, Xing Zhang, and Zhihao Cui for discussions and suggestions. 
\end{acknowledgments}

\section*{Data Availability Statement}

The data that support the findings of this study are openly
available on GitHub with the link: 

\href{https://github.com/sunchong137/data_for_nocisd_paper}{https://github.com/sunchong137/data\_for\_nocisd\_paper}

\bibliography{references}

\end{document}